\documentclass[alpha-refs]{wiley-article}

\usepackage{float}
\usepackage{siunitx}
\usepackage{hyperref}
\usepackage{dsfont}
\usepackage{bbm}
\usepackage{longtable}
\usepackage{algorithm}
\usepackage{algpseudocode}
\usepackage{pifont}
\usepackage[authormarkup=superscript,]{changes}
\definechangesauthor[name={TMOTTM},color=red]{HC}

\papertype{Original Article}
\paperfield{Journal Section}

\title{A Maximum Weighted Logrank Test in Detecting Crossing Hazards }

\author[1]{Huan Cheng}
\author[1]{Jianghua He}

\affil[1]{Department of Biostatistics and Data Science, University of Kansas Medical Center, Kansas City, KS, 66103, US}

\corraddress{Jianghua He}
\corremail{jhe@kumc.edu}

\begin{document}

\maketitle

\begin{abstract}
In practice, the logrank test is the most widely used method for testing the equality of survival distributions. It is the optimal method under the proportional hazard assumption. However, since non-proportional hazards are often encountered in oncology trials, alternative tests have been proposed. The maximum weighted logrank test was shown to be robust in general situations. In this manuscript, we propose a new maximum test that incorporates the weight for detecting crossing hazards. The new weight is a function of the crossing time-point. Extensive simulation studies are conducted to compare our methods with other methods proposed in the literature under scenarios with various hazard ratio patterns, sample sizes, censoring rates, and censoring patterns. For crossing hazards, the proposed test is shown to be the most powerful one with a known crossing time-point. It has a similar performance as the Maxcombo test in the misspecified crossing time-point scenario. Under other alternative situations, the new test remains comparatively powerful as the Maxcombo test. Finally, we illustrate the test in a real data example and discuss the procedures to extend the test to detect crossing hazards specifically.  

\end{abstract}


\section{Introduction}
In medical research, such as oncology studies, time-to-event outcomes are often used as clinical endpoints. It is of primary interest to compare the survival distributions between treatments and quantify the treatment effects. In practice, the classic logrank test is the most popular testing method, and Cox regression is used for estimating the treatment effects. Though it is well known that both logrank test and Cox regression are optimal under proportional hazards assumption, the research by \cite{jachno2019non} shows that the majority of the reviewed studies used the logrank test (88\%) and Cox regression (97\%)  for the primary outcome and only a few checked the proportional hazards assumption. Methodologies assuming proportional hazards were predominantly used despite the potential power improvements with alternative methods. 

The common types of alternatives to the proportional hazards include the delayed treatment effects, gradually increasing treatment benefits, diminishing treatment effects, and crossing treatment effects (i.e., initial adverse event and long-term benefits or vice versa). Several methods were proposed to improve the test power under non-proportional hazards assumptions.

The first type of test is based on weighted logrank test statistic. These tests give observed risk differences different weights at different time points. The famous Fleming-Harrington , also called $G^{\rho,\gamma}$ test (\cite{harrington1982class}, \cite{fleming2011counting}) is of this type. The weight is based on the Kaplan-Meier estimate of the pooled survival function at the previous event time.   A vast of literature studied the properties of the test and proposed different weights even before the Fleming-Harrington test. For example, \cite{gehan1965generalized} used the number at risk in the combined sample ($Y_i$) as weight and yielded the generalized two-sample Mann-Whitney-Wilcoxon test. \cite{tarone1977distribution} suggested a class of tests where the weight is a function of the number at risk ($f(Y_i)$). They suggested using $Y_i^{1/2}$, which gives more weight to the event time points with most data. See \cite{arboretti2018nonparametric}
 for a thorough summary of the weights and corresponding test names. The researchers should be cautious in choosing a proper weight function. If they have prior knowledge of the direction of the alternatives, a function that puts more weight on the departure of the hazards can be chosen to improve the power. Otherwise, an improper weight may perform worse than the logrank test. Adaptive tests were proposed to circumvent specifying weights beforehand. Essentially, these tests are also based on the weighted logrank statistic. The adaptive property reflects in the weight estimation and selection. \cite{peckova2003adaptive} proposed an adaptive test that selects a weight from a finite set of statistics based on efficiency criteria. Under the assumption of a time-transformed shift model,  the length of the confidence interval for the shift is used as an efficiency estimator for each test. The statistic with maximum efficiency is selected in the procedure. Although they suggested using $ln(t)$ as the transformation function, the specification has an impact on the power of the test. \cite{yang2005semiparametric} proposed a hazard model that accommodates different scenarios. The parameters in the model have the interpretations of short-term and long-term hazard ratios. An adaptive test (\cite{yang2010improved}) based on the weights estimated from the proposed model is proposed. The weight functions are hazard ratio estimates as $\Phi_1=\frac{\hat{\lambda}_0}{\hat{\lambda}_1}$ and $\Phi_2=\frac{\hat{\lambda}_1}{\hat{\lambda}_0}$. The corresponding test is defined as $\phi_{n,\alpha}=\boldsymbol{1}\{\text{max}(|W_{\Phi_1}|,|W_{\Phi_2}|)>
z_{\alpha/2}\}$. It is shown to be more powerful under various alternative cases, but the test has an overly inflated type I error according to \cite{chauvel2014tests}. Another issue we found is that the hazard estimate - $\hat{\lambda}_0,\hat{\lambda}_1$ are model-based and asymmetric. If the labels of the groups will be flipped, the test statistics and p-values are different. This is not a feature a test desires.  The maximum weighted logrank test that takes the maximum of a finite cluster of statistics can also be considered as an adaptive test in the sense that it does not require pre-specification of weight functions and automatically select the largest one.  \cite{lee1996some} proposed a test based on the maximum of selected members of $G^{\rho,\gamma}$ statistics. The weights addressing different types of alternatives are pooled together. \cite{gares2015omnibus} focused on the late treatment effect in the preventive randomized clinical trial. They proposed a maximum test based on the logrank statistic and several Fleming-Harrington statistics for late effect, which showed power improvements under late effect. \cite{lin2020alternative}
examined a number of tests under non-proportional cases and proposed the so-called MaxCombo test, a maximum test based on specific $G^{\rho,\gamma}$ statistics because of its robustness across different patterns of the various hazard ratios.  \cite{brendel2014weighted}  and\cite{ditzhaus2020more} proposed the projection test, which combines a cluster of statistics by mapping the multiple statistics into one single statistic. The power advantage over various methods was illustrated.

 The second one includes the Renyi-type or supremum statistic. It's the generalization of Kolmogorov-Smirnov statistic. The test statistic takes the maximum difference across the time points. \cite{fleming1981class}, \cite{fleming1987supremum} proposed the weighted version of the Renyi statistics, where the maximum is based on the weighted logrank statistics. Those supremum versions of logrank statistics are assumed to be more sensitive to cases with crossing hazard functions.

 The third type is based on the survival curves. \cite{pepe1989weighted} proposed the weighted Kaplan-Meier Statistic(WKM), which is based on the integrated weighted difference in Kaplan-Meier estimators. The test is sensitive against stochastic ordered alternatives. \cite{liu2020resampling} used the scaled area between curves(ABC) statistic, which is based on the absolute difference between the Kaplan-Meier estimators.

In this paper, we proposed a new type of maximum weighted logrank test. Overall, the maximum logrank tests are very robust against different types of hazards, as shown in the previous studies. However, the cluster of statistics is all based on the $G^{\rho,\gamma}$ statistics, according to our research. Depending on the selection of $\rho, \gamma$, the statistics are sensitive to early, middle, or late hazard differences. Therefore, we are motivated to introduce a type of weight that preserves the advantage and is more sensitive to crossing hazards. So the power of the maximum test can be further improved in detecting crossing hazard alternatives. 
We organize the paper as follows: In Section 2, we briefly review the weighted logrank test and methods used for later comparison; in Section 3, we introduce the new test method; in Section 4, the new test method is examined in simulation studies along with several comparative tests. Next, we illustrate the new method in a real data example in Section 5. Finally, a brief discussion and summary conclude the paper in Section 6.

\section{Background}
\subsection{Two-sample data set-up}
\begin{figure}[H]
    \centering
  \includegraphics[scale=0.7]{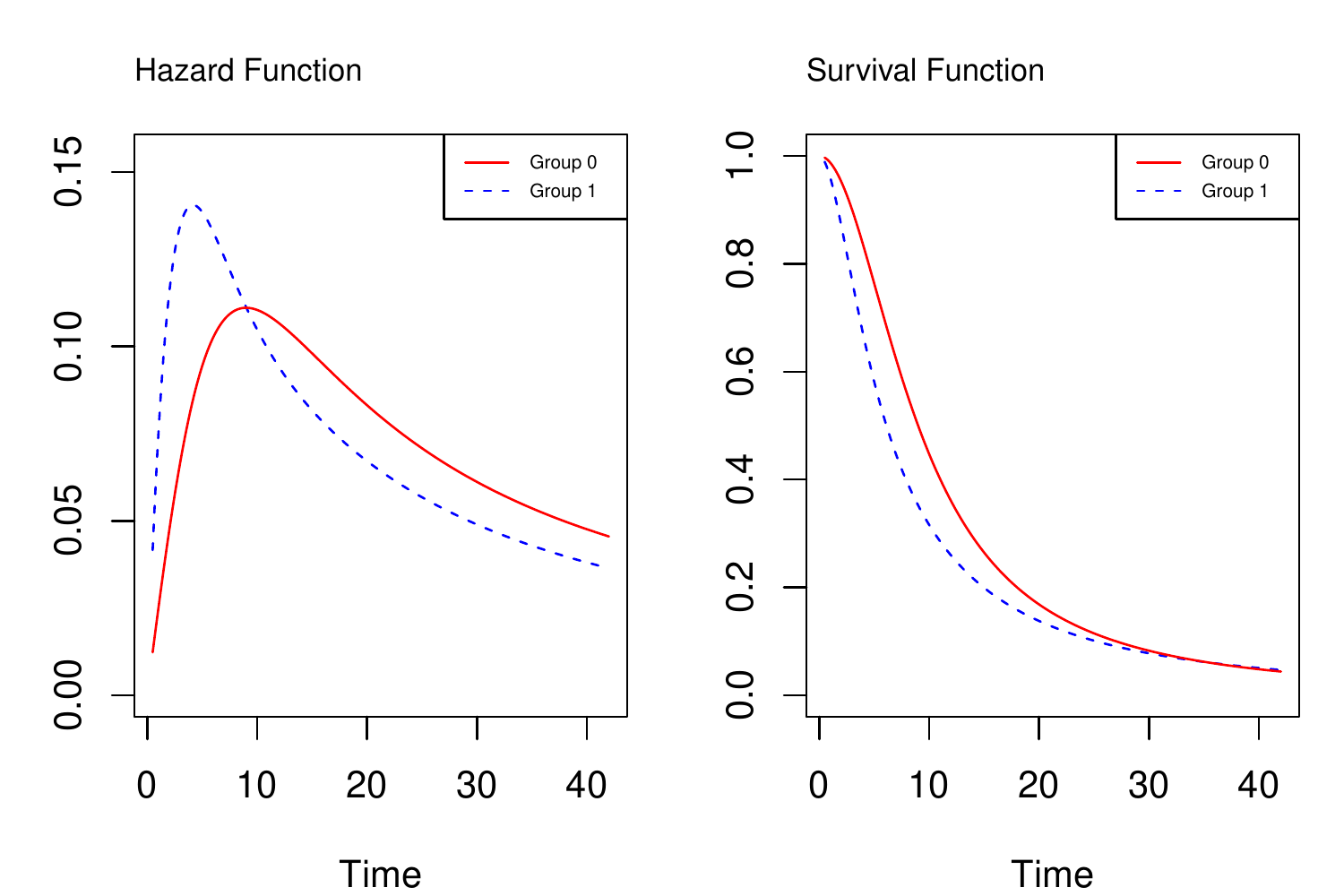}
    \caption{An example of stochastic hazard alternative }
    \label{fig:example1 }
\end{figure}

We focus on the standard setting with two-sample right censored survival data. The classic set-up is given by the event time $T_{ij} \sim F_i$ and censoring time $C_{ij} \sim G_i$, where $T_{ij}$ and $C_{ij}$ are the event time and censoring time of subjects $j\;  (j=1,...,n_i)$ from group $i \;(i=0,1)$ and $F_i$ and $G_i$ are the cumulative distribution functions. The event time $T_{ij}$ is assumed to be independent of censoring time $C_{ij}$. In practice, the available data only consists of $\big\{X_{ij}=T_{ij}\bigwedge C_{ij}$, $\delta_{ij}=\boldsymbol 1\{T_{ij}\leq C_{ij}\}\big\}$, where $X_{ij}$ denotes the observed time of subject $j$ in group $i$ and $\delta_{ij}$ indicates whether the observation is an event or censoring. 
Let $\Lambda_i(t)$ denote the cumulative hazard function and $\lambda_i$ denote the hazard rate. $F_i,\Lambda_i$ and $\lambda_i$ are related by $\Lambda_i(t)=-log(1-F_i(t))=\int_0^t\frac{dF_i}{1-F_i}=\int_0^t\lambda_i(s)ds$. The goal is to test 
\begin{equation*}
    H_0: \{F_0(s)=F_1(s)\}=\{\Lambda_0(s)=\Lambda_1(s)\},\; \text{for all } s \geq 0
\end{equation*}
against two sided alternatives $A_0:\{\Lambda_0(s)\neq\Lambda_1(s)\}$ for some $s$;  stochastic ordered alternatives $A_1:\{\Lambda_0(s) \geq \Lambda_1(s)\}$ where the inequality is strict for at least some $s$ or ordered hazard alternatives $A_2:\{\lambda_0(s) \geq \lambda_1(s)\}$ where the inequality is strict for at least some $s$.
Clearly, $A_2$ implies $A_1.$ Alternative $A_1$ includes the crossing hazards but neither $A_1$ nor $A_2$ consists of the case where survival curves cross. The example in Figure \ref{fig:example1 } illustrates the stochastic ordered alternatives. Though the survival rate of Group 1 is stochastically worse than Group 0, the hazard of Group 1 is greater at the beginning and less than the hazard of Group 0 in the longer term. If a two-sided test is performed, we should be cautious in the interpretation of the treatment effect. \cite{magirr2019modestly} discusses the risk of concluding the treatment is efficacious when it is uniformly inferior, as shown in Figure \ref{fig:example1 }. They proposed to use a "strong" null hypothesis that survival in the treatment group is stochastically less than or equal to survival in the control group and recommended consistently using one-sided hypothesis tests in confirmatory clinical trials. 

\subsection{Review of selected tests }
\subsubsection{Weighted logrank tests}
\cite{gill1980censoring} introduced a class of linear rank statistic and named ithem  as "tests of the class $\mathcal{K}$". This class $\mathcal{K}$ statistics are the so-called weighted logrank statistics. 
In the notation of counting process ( \cite{fleming2011counting}), the classic weighted logrank test can be expressed as
\begin{equation}
\label{wk}
    W_k(t)=\int_0^{t} K(s)\frac{dN_1}{Y_1}-\int_0^{t} K(s)\frac{dN_0}{Y_0}
\end{equation}
where 
\begin{equation*}
    \begin{split}
        K(s)&=(\frac{n_1+n_0}{n_1n_0})^{1/2}w(s)\frac{Y_1(s)Y_0(s)}{Y_1(s)+Y_0(s)}\\
    N_i(s)&=\sum_{j=1}^{n_i}I[X_{ij} \leq s,\delta_{ij}=1]\\
    Y_i(s)&=\sum_{j=1}^{n_i}I[X_{ij} \geq s]
    \end{split}
\end{equation*}
$w(s)$ is an adapted and bounded predictable process with $w(s)=0$ if $Y_1(s)\bigwedge Y_0(s)=0$.
$\int \frac{dN_i}{Y_i}$ is an estimator of the cumulative hazard $\Lambda_i(t)$. 
Thus, $W_k$ is essentially a sum of weighted differences in the estimated hazards over times. Equation (\ref{wk}) can be rewritten as 
\begin{equation}
\label{WLRT2}
    W_k(t)=\int_0^{t} K(s)d\hat{\Lambda}_1(s)-\int_0^{t} K(s)d\hat{\Lambda}_0(s)
\end{equation}
Under $H_0$, $\{W_k(t),0\leq t\leq \infty\}$ is asymptotically distributed as  a zero-mean Gaussian process with variance function (\cite{gill1980censoring}, \cite{fleming2011counting})
\begin{equation*}
\sigma^2_n(W_k)=    \int_0^t w(s)^2\frac{P(X_1 \geq s)P(X_0 \geq s)}{\eta P(X_1 \geq s) +(1-\eta )P(X_0 \geq s)}(1-\Delta \Lambda(s))d\Lambda(s)
\end{equation*}
where $\lim \frac{n_1}{n_1+n_0}=\eta \in [0,1]$. 
An estimator of $\sigma^2$ is given by 
\begin{equation}
\label{variance}
\hat{\sigma}^2_n(W_k)=    \frac{n_1+n_0}{n_1n_0}\int w(s)^2\frac{Y_1Y_0}{Y_1+Y_0}(1-\frac{\Delta N_1+\Delta N_0-1}{Y_1+Y_0-1})\frac{d(N_1+N_0)}{Y_1+Y_0}
\end{equation}
The $G^{\rho,\gamma}$ class statistic is given by $w(s)=\{\hat{S}(s)\}^{\rho}\{1-\hat{S}(s)\}^{\gamma},\rho \geq 0,\gamma \geq 0$, where $\hat{S}(s)$ is the left continuous Kaplan-Meier estimator of survival function based on pooled data.  $\{\gamma=0$, $\rho=0\}$ corresponds to the logrank statistic. $\{\gamma=0,\rho=1\} $ corresponds to the Prentice-Wilcoxon statistic, which emphasizes early difference. It's shown that the logrank test is most efficient under proportional hazard and $G^\rho, \rho>0$ statistic (Prentice-Wilcoxon test is a special case) is efficient against monotone decreasing hazard ratio alternative. (\cite{fleming2011counting}).
In general, a test with weight proportional to $\log(\lambda_1(s)/\lambda_0(s))$ is most powerful under alternatives (see \cite{schoenfeld1981asymptotic}). In practice, if a prior knowledge about the direction of the hazards can be obtained, a proper weight can be chosen to emphasize the difference. For example, if there is evidence to believe the two groups have maximum difference in the middle of the follow-up time period, weight $\{\gamma=1,\rho=1\} $ gives more power. However, if the underlying hazards are diverging over time, an erroneous chosen statistic for example, $\{\gamma=0,\rho=1\} $ performs worse than the logrank test. 
\subsubsection{Renyi test }

The Renyi test statistic is the censored-data version of the Kolmogorov-Smirnov statistic for uncensored samples. To construct the test, we should find the value of the test statistic (\ref{WLRT2}) at each event time point. The test statistic for a two-sided test is given by 
\begin{equation*}
    Q=\text{sup}\{|W_k(s)|,\;s\leq t\}/\sigma(t)
\end{equation*}
where $t$ is the largest event time with $Y_1(t)\bigwedge Y_0(t)>0$. Under null hypothesis, the distribution Q is approximated by the distribution of $\text{sup}\{|B(x)|,\;0\leq x\leq1\}$, $B$ is a standard Brownian motion process (\cite{klein2006survival}). This superemum version of the test statistics is more powerful in detecting crossing hazards. 
The weight function $w(s)$ introduced above can be used for Renyi test. Accordingly, we obtain various versions of Renyi test by choosing different weights. 

\subsubsection{Projection-type test }
\cite{brendel2014weighted} introduced the projection-type test and showed the asymptotic proprieties of the statistic. \cite{ditzhaus2020more} further clarified and simplified the methods. The way to construct the test statistic is as follows: choose different weights $w_1,...,w_m$ representing different directions for the hazard and construct the the vector $\boldsymbol{T_n}=[T_n(w_1),...,T_n(w_m)]$, where $T_n(w_i)$ is the $W_k(t)$ given by (\ref{WLRT2}) with weight $w_i$ . In the paper of \cite{ditzhaus2020more}, they consider linearly independent weights. If the independence assumption is met, under the null hypothesis, we have
\begin{equation*}
    S_n=T_n^T\hat{\Sigma}^{-1}_nT_n \xrightarrow{D}\chi_m^2
\end{equation*}
where $\hat{\Sigma}_n$ is the empirical covariance matrix of $T_n$. The corresponding test is given by  $\phi_{n,\alpha}=\boldsymbol 1\{S_n >\chi_{m,\alpha}^2\}.$ In the paper of \cite{brendel2014weighted} , independent assumption is not needed. The test statistic is given by $S_n=T_n^T\hat{\Sigma}^{-}_nT_n \xrightarrow{D}\chi_k^2,\text{where}\; k=rank(\hat{\Sigma}^{-}_n)$, where $\Sigma^{-}$ stands for the Moore-Penrose inverse of the covariance matrix. In the later comparison, the projection test refers to this one. 
\section{Proposed Test}
In this section, we constructed a new test statistic, that has good power against varied alternatives and is especially sensitive in detecting crossing hazards. The statistic is based on a number of statistics in the form of equation (\ref{wk}). In what follows, we will present the asymptotic distributions and procedures to obtain the p-value.

Let $W_{k_1},...,W_{k_m}$ denote the statistic as described in (\ref{wk}). The weight function for each statistic can be expressed as $W_i(s)=\Tilde{w}_i\circ \hat{F}_n(s-)$,  where $\Tilde{w}_i$ is a deterministic function and $\hat{F}_n(s-)$ is the pooled Kaplan-Meier estimate of CDF. For instance, the Prentice-wilcoxon statistic corresponds to function $\Tilde{w}(u)=1-u$. For those $m$ statistics, the difference lies in the choice of the deterministic functions.  Under $H_0:F_1=F_0$, we have 
\begin{equation}
\label{convg}
    (W_{k_1},...,W_{k_m})\xrightarrow[]{D}(Z_1,...,Z_m)
\end{equation}
where $(Z_1,...,Z_m)$ is a mean zero m-variate Gaussian random vector. The covariance between $Z_k$ and $Z_l$ is given by 

\begin{equation*}
\sigma_{k,l}(t)=    \int_0^t W_k(s)W_l(s)\frac{P(X_1 \geq s)P(X_0 \geq s)}{\eta P(X_1 \geq s) +(1-\eta )P(X_0 \geq s)}(1-\Delta \Lambda(s))d\Lambda(s)
\end{equation*}
A consistent estimator of $\sigma_{k,l}(t)$ is 
\begin{equation}
\label{covariance}
\hat{\sigma}_{k,l}(t)=    \frac{n_1+n_0}{n_1n_0}\int_0^t W_k(s)W_l(s)\frac{Y_1Y_0}{Y_1+Y_0}(1-\frac{\Delta N_1+\Delta N_0-1}{Y_1+Y_0-1})\frac{d(N_1+N_0)}{Y_1+Y_0}
\end{equation}
The two-sided maximum weighted logrank test statistic is defined as $\text{T}=\text{max}(|W_{k_1}|,...,|W_{k_m}|)$. Some researchers investigated the performance of maximum tests using $G^{\rho,\gamma}$ type weights. For example, \cite{lee1996some} proposed $\Tilde{w}_1=1,\Tilde{w}_2=u^2,\Tilde{w}_3=(1-u)^2,\Tilde{w}_4=(1-u)^2u^2,\; u \in (0,1)$. The four weights emphasize proportional hazards, late, early and middle difference in hazards. The test demonstrated power improvement under various hazards scenarios. \cite{kosorok1999versatility} used $G^{0,0}, G^{0,1}, G^{4,0}, G^{4,1}$, corresponding to $\Tilde{w}_1=1,\Tilde{w}_2=u,\Tilde{w}_3=(1-u)^4,\Tilde{w}_4=(1-u)^4u,\; u \in (0,1)$. Recently, \cite{lin2020alternative} used $G^{0,0}, G^{0,1}, G^{1,0}, G^{1,1}$ and named the test as Maxcombo test. Those methods focus on non-crossing hazards alternatives. Under crossing hazards and especially crossing survival curves, the weighted logrank tests usually perform poorly. Conceptually, it is not difficult to understand because the weighted logrank statistic is a sum of differences in hazards over time. Under ordered hazards alternatives, the signs of the summation components are stochastically in the same direction. Correctly putting more weights where the differences are large improves the test power. However, in terms of crossing hazards, as shown in Figure \ref{fig:example1 }, the summation components tend to have opposite signs before and after the crossing time point. Whether the emphasis is put on the beginning, middle, or end times, similar to the logrank test, the early differences between two hazard rates are canceled out by later differences with opposite signs. The Maxcombo test introduced previously may improve the power to detect crossing hazards, but there is still room for improvement. 

\begin{figure}
    \centering
    \includegraphics[scale=0.4]{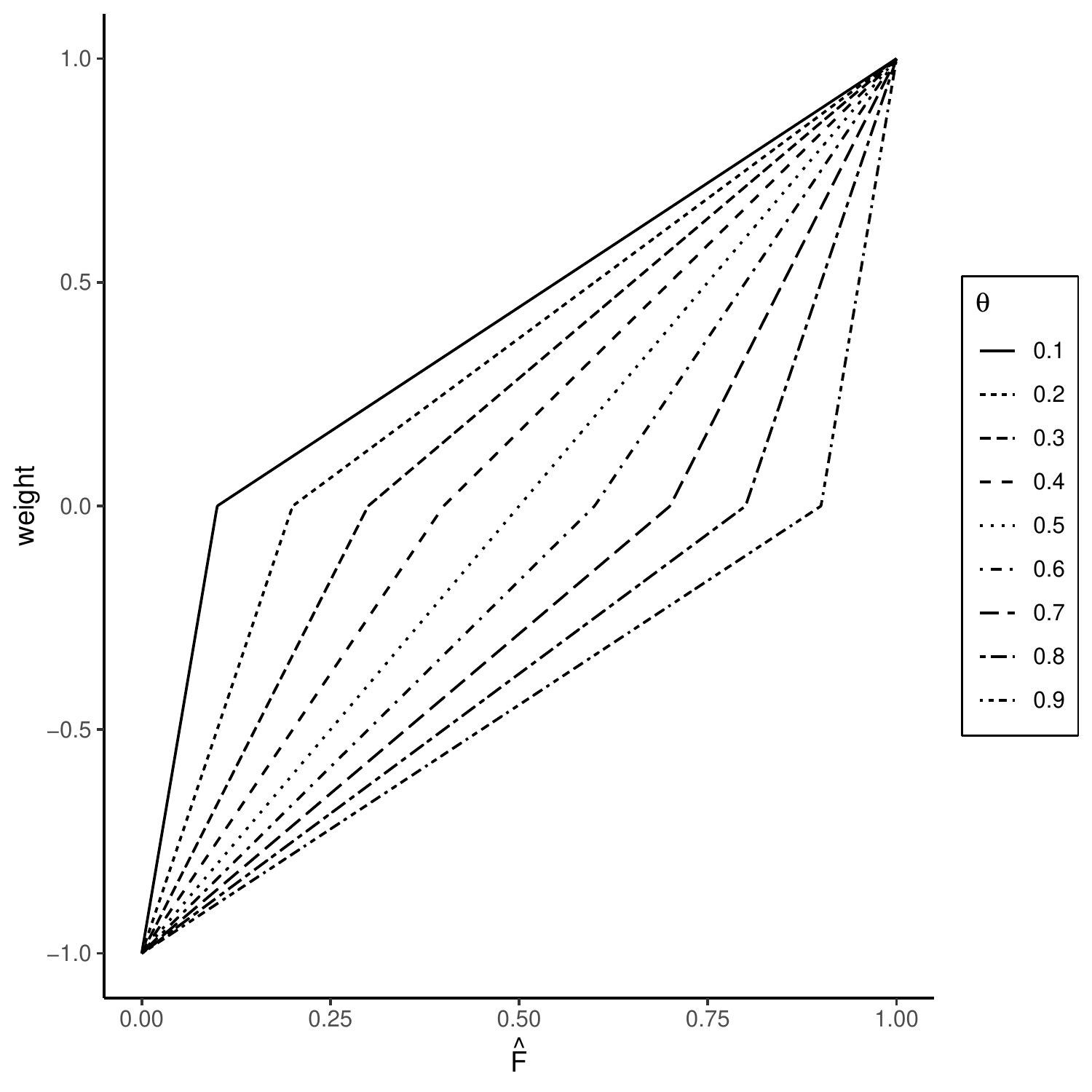}
    \caption{Crossing weights with different $\theta$s}
    \label{fig:demo_wgt}
\end{figure}

Our proposed test is to improve the power in detecting crossing hazards and maintain good power in other alternatives as well. The four weights in the new test are - $\Tilde{w}_1=1,\Tilde{w}_2=u,\Tilde{w}_3=1-u,\Tilde{w}_4=\mathbb{1}(u\leq \theta)\frac{u-\theta}{\theta}+\mathbb{1}(u> \theta)\frac{u-\theta}{1-\theta}$, where $u$ is the pooled Kaplan-Meier estimate of  CDF and $\theta \in (0,1)$ represents the value of $u$ at the point where the crossing occurs. In practice, $\theta$ is unknown. So we recommend the default value 0.5, simplifying $\Tilde{w}_4$ to $2u-1$. Figure \ref{fig:demo_wgt} shows $\Tilde{w}_4$ with $\theta$ varying from 0.1 to 0.9. As the CDF increases with time, a smaller value for $\theta$ is preferable if the researchers believe the crossing occurs at an early stage; otherwise, a larger value is recommended. As shown in the simulation, our test is robust to the choice of $\theta$. Even if a smaller $\theta$ is chosen for a late crossing, the proposed test still outperforms the logrank test and has similar performance to Maxcombo test. 

For a two-sided test, the rejection region and p-value of the proposed test are obtained based on the asymptotic normality. According to (\ref{convg}), $(Z_1,...,Z_m)$ follows a multivariate normal distribution with estimated variance-covariance given by (\ref{variance}, \ref{covariance}). Let $c_\alpha$ denote the critical value under significance level $\alpha$. To control the type I error under $\alpha$, we have 
$P(T\geq c_\alpha)=\alpha$, equivalent to $P(|W_{k_1}|<c_\alpha,|W_{k_2}|<c_\alpha,|W_{k_3}|<c_\alpha) =1-\alpha$. Here $c_\alpha $ is obtained via finding the corresponding critical values of the multivariate normal distribution. In the extreme case that the four statistics are independent, the equation above is simplified as  $P(|W_{k_1}|<c_\alpha) =(1-\alpha)^{1/4}$. So $c_\alpha=\Phi(\frac{1}{2}+\frac{(1-\alpha)^{1/4}}{2})$. In the opposite case when the four statistics are the same, we have $c_\alpha=\Phi(1-\frac{\alpha}{2})$. In general, the critical value $c_\alpha$ ranges between $\Phi(1-\frac{\alpha}{2})$  and $\Phi(\frac{1}{2}+\frac{(1-\alpha)^{1/4}}{2})$. The p-value is easy to obtain accordingly. 

For a one-sided test, the four-weight test statistic becomes $T=sign(W_{k_1})\times max(|W_{k_1}|,|W_{k_2}|,|W_{k_3}|,|W_{k_4}|)$. Consider the ordered alternative $H_1:\Lambda_0 \geq \Lambda_1$,where the strict inequality holds for some time, the type I error is defined as $P(T\leq c_\alpha)=\alpha $; for the alternative $\Lambda_1 \geq \Lambda_0$, the type I error is given by $P(T\geq c_\alpha)=\alpha $

\section{Simulation Study}
\subsection{Simulation Settings}
Two censoring mechanisms are considered in this simulation.
Under censoring Type I, the length of the study is fixed, and the number of events depends on the study duration and hazard rate. A complete study includes an 18-week recruitment period and is terminated at week 42. Every participant has at least 24 weeks of follow-up if no event occurs. Under censoring Type II, the study is terminated once the specified number of events is obtained. The overall event rate is fixed, but the total length of study depends on the enrollment and hazard rates. We assume the participants are uniformly enrolled within 24 weeks. In both settings, the end of the study is the only reason for censoring. 

\begin{figure}
    \centering
    \includegraphics[scale=0.5]{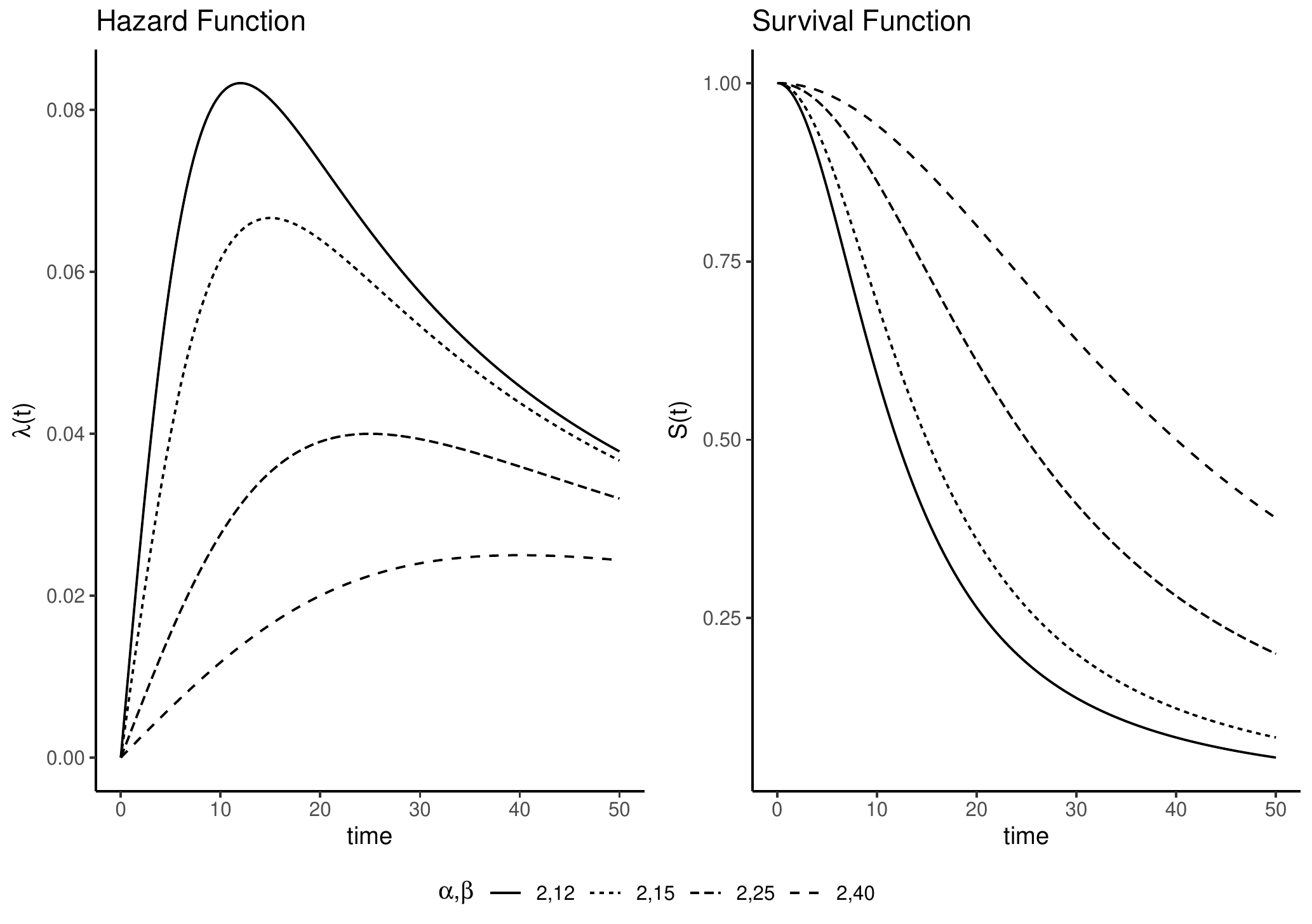}
    \caption{Hazard and survival function based on loglogistic distribution }
    \label{fig:demo_st_ht}
\end{figure}

Survival times for Group 0 are drawn from a log-logistic distribution with shape parameter $\alpha$ and scale parameter $\beta $. The hazard function is $\lambda_0(t)=\frac{\alpha/\beta (t/\beta)^{\alpha-1}}{1+(t/\beta)^\alpha}$ and the survival function is $S(t)=\frac{\beta^\alpha}{\beta^\alpha+t^\alpha }$. Figure \ref{fig:demo_st_ht} shows the hazard functions and survival functions plotted for a selection of parameters -$\alpha,\beta $. The shape parameter $\alpha $ is fixed at 2. While scale parameter $\beta$ increases from 12 to 40, the sharp peak of the hazard curve gets flattened and the survival curve is pulled to the upper right. The hazard function for the Group 1 is assumed to be a multiplier of the hazard of Group 0, that is, -$\lambda_1(t)=g(t)\lambda_0(t)$, where $g(t)$ is the hazard ratio between the two groups. In the proportional hazard case, $g(t)$ is a constant of time. To cover a wide range of alternatives, $g(t)$ has the following eight options. 
\begin{enumerate}[label=(\Alph*)]
 \item Crossing Hazards 1:  $g(t)=\{0.5\mathbb{1}[t<10]+(\frac{t-10}{15}+0.5)\mathbb{1}[10\leq t\leq 25]+1.5\mathbb{1}[t>25]\}$
  
     \item Crossing Hazards 2:  $g(t)=\{3exp(-0.3t)+0.8\}$
     
 \item Delayed Diverging Hazards:
 $g(t)=\{\frac{1.5}{1+exp(-0.5(t-20))}+1\}$

    \item Diverging Hazards: $g(t)=exp(0.03t)$
 
    \item Converging Hazards 1:  $g(t)=exp(\frac{1}{0.2t+1})$
  
     \item Converging Hazards 2:  $g(t)=\{(1-(t-50)^2/5000)\mathbb{1}[t\leq 40]+0.98\mathbb{1}[t>40])\}$
     
    \item Constant Hazards: $g(t)=1.5$
    \item Equal Hazards ($H_0$): $g(t)=1$
\end{enumerate}
The hazard ratios are time-dependent in case (A) - Case (F). The cumulative hazard for Group 1 is  $\Lambda_1(t)=\int_0^tg(u)\lambda_0(u)du$. Survival function is $S_1(t)=\text{exp}(-\Lambda_1(t))=\text{exp}(-\int_0^tg(u)\lambda_0(u)du)$. it is known that $S_1 \sim \text{uniform}(0,1)$. The survival times for Group 1 are generated via the inverse method, that is, $t_1=S_1^{-1}(u), \quad u \sim \text{uniform}(0,1)$. 

\begin{figure}
    \centering
   \includegraphics[scale=0.8]{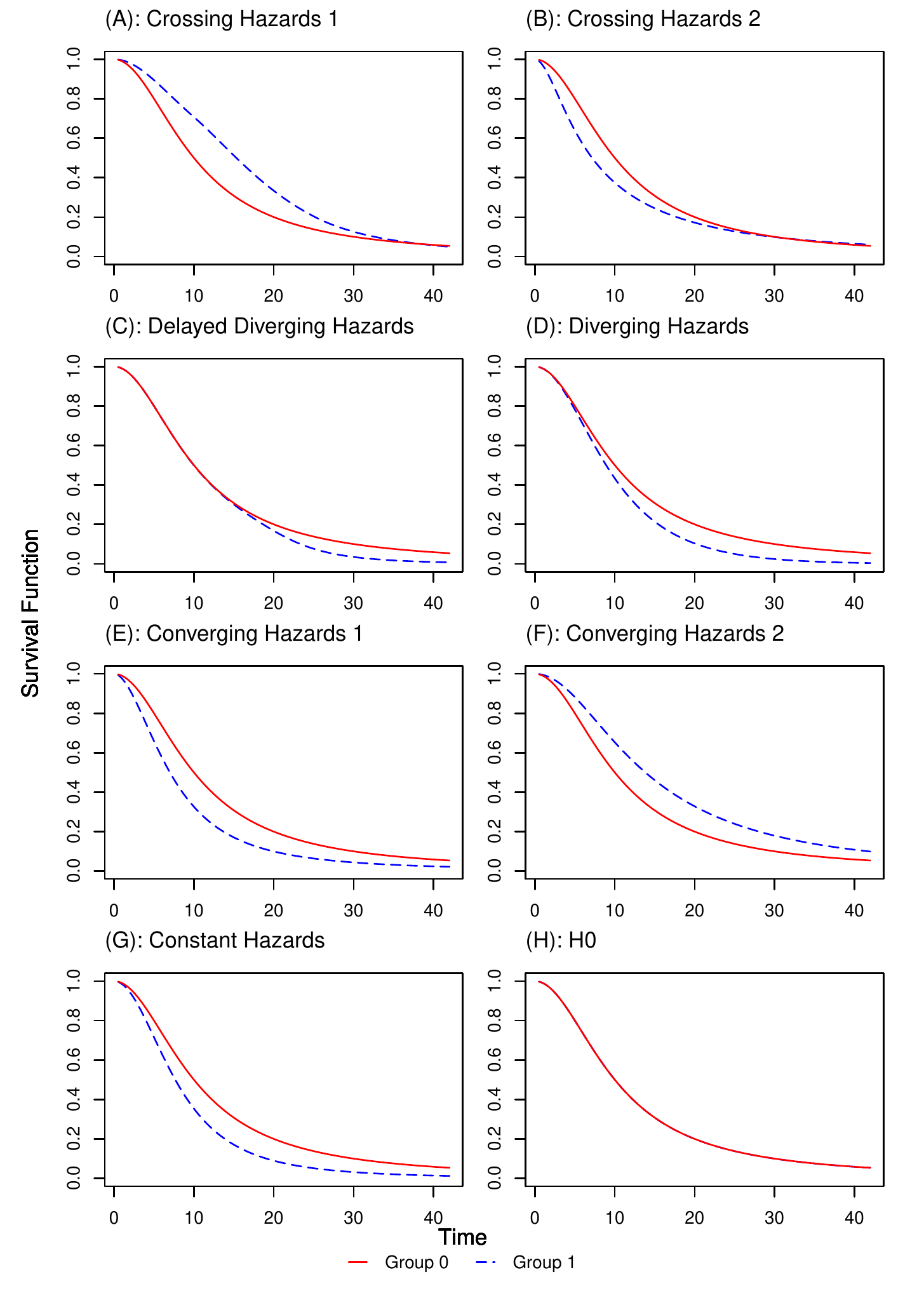}
    \caption{Survival Functions ($\alpha=2,\beta=15$)}
    \label{fig:demo_st}
\end{figure}

\begin{figure}
    \centering
   \includegraphics[scale=0.8]{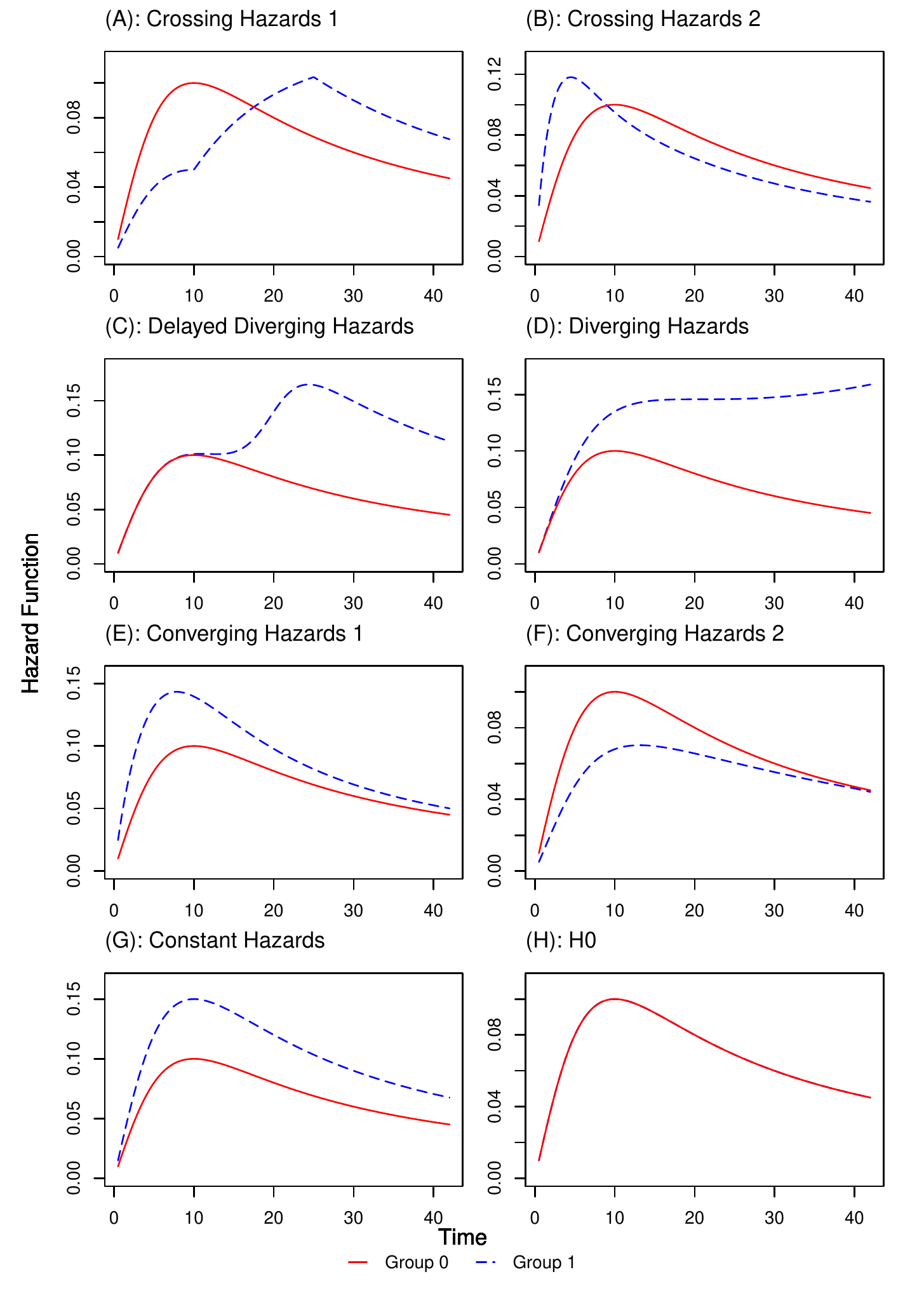}
    \caption{Hazard Functions ($\alpha=2,\beta=15$)}
    \label{fig:demo_ht}
\end{figure}

The survival curves and hazard curves of the eight different cases are given in Figure \ref{fig:demo_st} and Figure \ref{fig:demo_ht} ($\alpha=2,\beta=15$ for Group 0). Case (A) and case (B) represent crossing hazards. In (A), hazard ratio ($\lambda_1/\lambda_0$) is less than one at the beginning, then the two hazard curves cross around week 20, and the survival curves cross around week 35. In case (B), it is the other way around, that is, hazard ratio is greater than one at the beginning. Cases (C) and (D) show the diverging hazards of the two groups. If the hazard ratio is greater than one, the ratio increases over time; otherwise, it decreases over time. In the simulation, we consider the former case. In case (C) it illustrates the delayed response, where the medication takes effect after a certain amount of time. This is common in cancer vaccine trial (\cite{copier2009improving}). Case (E) and case (F) both show the converging hazards over time. In case (E), the hazard ratio decreases to close to 1, so the survival curve of Group 1 is below the curve of Group 0.  In case (F), the hazard ratio increases to close to 1, and Group 1 survives longer. Case (G) is for the proportional hazards with a ratio of 1.5. Case (G) denotes the null hypothesis where the two groups have no difference.

To investigate the operating characteristics of the proposed method, a variety of sample sizes and censoring rates are considered. Under censoring Type I, the study length is predefined. The censoring rate $\phi$ is altered by changing the parameters of the survival distribution. The shape parameter $\alpha$ is set to 2 through simulation, while the scale parameter - $\beta$ is set to $15,25,40$, corresponding to the low, medium, and high censoring rate. Under censoring Type II, sample size $N=60, 120, 240$ and censoring rate $\phi=1/6,1/3,1/2$ are assumed, yielding nine different combinations of $N$ and $\phi$. The specified censoring rate is maintained by changing the study length to ensure the specific number of events is obtained. The parameters for generating the Group 0 survival times remain the same across all the combinations with $\alpha=2,\beta=12$. For both mechanisms, the sample size is allocated to the two groups at a 1:1 ratio. 

We mainly consider the two-sided test in the above simulation settings and the one-sided test under the stochastic ordered alternative. The test level $\alpha$ is $0.05$ for the two-sided test and $0.025$ for the one-sided test. All simulations are run in R, version 3.5.3 platform: x86\_64-pc-Linux-gnu with 2000 replications. The largest simulation error is less than 0.01.


\subsection{Results}
Six tests are considered in the two-sided test simulations. They are the logrank test, Fleming-Harrington test with $\gamma=1,\rho=1$, the MaxCombo test with four weight functions - $G^{0,0}, G^{0,1}, G^{1,0}, G^{1,1}$, the proposed test with $\theta=0.5$, the projection test and Renyi's test. To account for the crossing hazards, the weights used for the projection test are $\{1,u,2u-1\}$. The six tests are denoted by Logrank, FH11, maxC, $\varphi^*(0.5)$, ProjT and Renyi in the following tables. Considering the overly inflated Type I error in Yang's method(\cite{yang2010improved}) and the moderate performance of the weighted Kaplan-Meier method in the study of \cite{lin2020alternative}, we did not include the two methods for comparison. 

Under the null hypothesis that the two groups are not different, the two censoring mechanisms generate almost the same censoring rates at each group. The minor difference is simply due to the sampling variation. The Type I error rates for the Type I censoring and Type II censoring are displayed in Table \ref{tab:fres1} and Table \ref{tab:res1}. Except for Renyi's test, all the remaining tests have inflated Type I error at sample size=60 regardless of censoring mechanism or censoring rate, although increasing censoring rate brings the Type I error rate down. At sample size=240, most of the tests control the Type I error rate under 0.05. 



\begin{longtable}[r]{p{0.1\textwidth}p{0.04\textwidth}p{0.04\textwidth}p{0.06\textwidth}p{0.06\textwidth}p{0.06\textwidth}p{0.06\textwidth}p{0.06\textwidth}p{0.06\textwidth}}
\caption{Type I error rates under Type I censoring mechanism (\%)}
\label{tab:fres1}\\
  \hline
   N & $\phi_1$ & $\phi_0$ & Logrank & FH11 & maxC & $\varphi^*(0.5)$  & ProjT & Renyi \\ 
  \hline
  \endhead
  \hline   
  \endfoot

  60 & 0.18 & 0.18 &  6.5 & 6.0 & 6.3 & 6.5 & 6.4 & 5.6 \\ 
 
120 & 0.18 & 0.18 &  6.4 & 6.2 & 6.4 & 5.9 & 6.0 & 5.6 \\ 

240 & 0.18 & 0.18 &  5.3 & 5.4 & 5.6 & 5.2 & 4.8 & 5.2 \\ 

60 & 0.37 & 0.38 &  6.6 & 5.8 & 6.8 & 6.6 &  5.8 & 5.2 \\ 

120 & 0.37 & 0.38 & 5.4 & 5.9 & 5.8 & 5.8 &  5.7 & 4.6 \\ 

240 & 0.37 & 0.37 &  4.6 & 5.0 & 4.8 & 4.0 &4.0 & 4.2 \\ 

60 & 0.59 & 0.60 &  5.4 & 5.4 & 5.8 & 5.7 &  5.0 & 4.2 \\ 

120 & 0.60 & 0.60 &  5.2 & 4.4 & 5.1 & 4.9 &  4.9 & 4.2 \\ 
240 & 0.60 & 0.60 &  5.0 & 4.8 & 4.6 & 4.4 &  4.8 & 4.3 \\

   \hline
\end{longtable}


\begin{longtable}[r]{p{0.1\textwidth}p{0.04\textwidth}p{0.04\textwidth}p{0.06\textwidth}p{0.06\textwidth}p{0.06\textwidth}p{0.06\textwidth}p{0.06\textwidth}p{0.06\textwidth}}
\caption{Type I error rates under Type II censoring mechanism (\%)}
\label{tab:res1}\\
  \hline
  Event/N & $\phi_1$ & $\phi_0$ & Logrank & FH11 & maxC & $\varphi^*(0.5)$  & ProjT & Renyi \\ 
  \hline
  \endhead
  \hline   
  \endfoot
 
 50/60 & 0.17 & 0.17 & 6.6 & 6.2 & 7.0 & 6.4 & 6.5 & 5.2 \\ 
 100/120 & 0.17 & 0.17 & 6.6 & 6.1 & 6.6 & 6.1  & 6.1 & 5.6 \\ 
 200/240 & 0.17 & 0.17 & 6.0 & 5.6 & 5.2 & 4.8  & 4.4 & 5.4 \\ 
 40/60 & 0.33 & 0.34 & 5.8 & 5.2 & 6.1 & 6.0  & 6.2 & 4.4 \\ 
 80/120 & 0.33 & 0.34 & 6.1 & 6.1 & 6.5 & 6.0 & 5.8 & 5.0 \\ 
 160/240 & 0.33 & 0.33 & 5.2 & 4.8 & 5.2 & 4.7 & 4.8 & 3.8 \\ 
   30/60 & 0.50 & 0.50 & 5.9 & 5.6 & 6.2 & 5.8 & 5.6 & 5.0 \\ 
  60/120 & 0.50 & 0.50 & 5.5 & 5.4 & 6.0 & 5.6 & 5.2 & 5.0 \\ 
  120/240 & 0.50 & 0.50 & 4.6 & 5.0 & 4.8 & 4.4  & 4.7 & 4.4 \\ 
   \hline
\end{longtable}

Table \ref{tab:fres2} shows the empirical power of each test under the Type I censoring mechanism with the total study length of 42 weeks. The scale parameter $\beta$ for the Group 0 is fixed across different sample sizes giving the same censoring rates. The censoring rates (column $\phi_0$) for Group 0 are about 0.18,0.38 and 0.6, corresponding to 15, 25 and 40 of $\beta$. The average censoring rates for Group 1 are shown in column $\phi_1$ and the total sample sizes are given in column $N$. The hazard curve and survival curve will alter accordingly with the changing of the scale parameter. For the crossing hazards with $\phi_0=0.18$, the proposed test and projection test show a sizeable gain in power comparing to other tests. For sample size 240 and Crossing 1 alternative, both the proposed test and projection test have power over 80\%, while the logrank test and the Maxcombo test have the power of 26.6\% and 53.2\% respectively. When $\phi_0$ is about 0.38 or 0.6, the proposed test is not as powerful as the projection test, though it still outperforms all other tests, including the Maxcombo test. This is because the choice of $\theta=1/2$ is not reflecting the true CDF value at the crossing time point. The sensitivity of the proposed test to the choice of $\theta$ is discussed in the next section. Under diverging hazards and delayed response, the Maxcombo test has the largest power in most cases, followed by the proposed test. Under the converging and constant hazards, the logrank test is the most powerful one in most cases, and the projection test loses noticeable power. It is only more powerful than Renyi's test in some cases.


The power of the tests under various alternatives from the Type II censoring is shown in Table \ref{tab:res2_pwr}. The total event rates in the first column are pre-specified for each simulation. The average censoring rates for each group are given in the second and third columns denoted by $\phi_1$ and $\phi_0$.
The property of the design determines that it takes a longer follow-up time to obtain a larger number of events. In other words, the lower the censoring rate, the longer the follow-up time. The shapes of the two survival curves are the same under different censoring rates. However, the study termination times are different. This is different from the Type I censoring mechanism, where the study termination times are fixed at 42 weeks, but the shape of survival curves varies with censoring rates. At the crossing hazards, the proposed test and projection test have the largest power, showing a power advantage over the Maxcombo test. As expected, the study termination times have a significant impact on the power of the logrank test. For example, at a lower censoring rate - $\phi_1=18\%$ and $N=60$, the power of logrank test is 19\% under Crossing 1, while at a greater censoring rate $\phi_1=56\%$ and $N=60$, the power becomes 26.1\%, a sizable increase rather than decrease. Under ordered hazard alternatives, the increase in censoring rates brings the power of the test down. However, under the crossing hazard, the termination times play a more critical role in determining the power of the logrank test. The proposed $\varphi^*(0.5)$ test is more robust to the study termination point. In the same example, the power of the proposed test is 30.8\% at a lower censoring rate and 27.2\% at a large censoring rate, a slight decrease. At the diverging hazards, the Maxcombo test has the largest power in most cases, and the proposed test and projection test have comparable results. At the delayed response, the projection test and proposed test are the most powerful ones, followed by Maxcombo and logrank test. At the converging and proportional hazards, the logrank test is the most powerful one, the projection test and Renyi's test are the least powerful ones. The Maxcombo and proposed test are in between.

\begin{longtable}[r]{p{0.06\textwidth}p{0.04\textwidth}p{0.04\textwidth}p{0.16\textwidth}p{0.06\textwidth}p{0.06\textwidth}p{0.06\textwidth}p{0.06\textwidth}p{0.06\textwidth}p{0.06\textwidth}}
\caption{Empirical power under Type I censoring mechanism (\%)}
\label{tab:fres2}\\
  \hline
  N & $\phi_1$ & $\phi_0$ & Type & Logrank & FH11 & maxC & $\varphi^*(0.5)$  & ProjT & Renyi \\ 
  \hline
  \endhead
  \hline   
  \endfoot
60 & 0.18 & 0.18 & Crossing 1 & 11.2 & 7.5 & 15.2 & 25.0 & 25.9 & 9.8 \\ 
  60 & 0.20 & 0.18 & Crossing 2 & 6.8 & 5.7 & 9.8 & 13.4 & 14.9 & 5.4 \\ 
  60 & 0.07 & 0.18 & Delayed Diverging & 16.2 & 16.0 & 21.2 & 21.6  & 23.6 & 12.4 \\ 
  60 & 0.06 & 0.18 & Diverging & 31.7 & 31.9 & 35.2 & 32.4  & 32.4 & 25.8 \\ 
  60 & 0.10 & 0.18 & Converging 1 & 26.4 & 18.4 & 24.6 & 23.2  & 21.9 & 18.3 \\ 
  60 & 0.26 & 0.18 & Converging 2 & 18.4 & 12.6 & 17.6 & 18.0  & 16.2 & 12.2 \\ 
  60 & 0.08 & 0.18 & Constant & 32.8 & 29.0 & 31.2 & 27.8 & 26.6 & 26.0 \\ 
  120 & 0.18 & 0.18 & Crossing 1 & 16.5 & 8.3 & 29.0 & 47.8  & 49.0 & 14.4 \\ 
   120 & 0.20 & 0.18 & Crossing 2 & 7.8 & 6.2 & 13.2 & 23.0  & 24.6 & 5.9 \\ 
   120 & 0.07 & 0.18 & Delayed Diverging & 26.2 & 25.4 & 38.0 & 39.6 & 42.0 & 19.2 \\ 
120 & 0.06 & 0.18 & Diverging & 55.6 & 55.0 & 60.0 & 56.8  & 56.6 & 46.2 \\ 
120 & 0.10 & 0.18 & Converging 1 & 42.9 & 31.6 & 41.8 & 39.6  & 37.6 & 31.0 \\ 
120 & 0.26 & 0.18 & Converging 2 & 32.6 & 22.5 & 30.8 & 30.8  & 28.3 & 22.9 \\ 
120 & 0.08 & 0.18 & Constant & 55.8 & 50.5 & 52.4 & 47.2  & 44.2 & 46.6 \\ 
240 & 0.18 & 0.18 & Crossing 1 & 26.6 & 11.6 & 53.2 & 81.0  & 81.0 & 31.2 \\ 
240 & 0.19 & 0.18 & Crossing 2 & 7.6 & 6.2 & 20.9 & 43.0 & 45.8 & 6.3 \\ 
240 & 0.07 & 0.18 & Delayed Diverging & 46.6 & 43.6 & 67.9 & 70.1 & 73.3 & 34.6 \\ 
240 & 0.05 & 0.18 & Diverging & 84.4 & 83.0 & 89.2 & 87.0 & 87.0 & 76.4 \\ 
240 & 0.10 & 0.18 & Converging 1 & 70.6 & 55.5 & 70.6 & 68.8 & 67.1 & 55.8 \\ 
240 & 0.26 & 0.18 & Converging 2 & 54.4 & 39.8 & 55.5 & 55.8  & 53.1 & 41.2 \\ 
240 & 0.08 & 0.18 & Constant & 83.7 & 79.2 & 80.9 & 77.8  & 76.0 & 75.0 \\ 
60 & 0.35 & 0.38 & Crossing 1 & 5.9 & 8.0 & 9.8 & 13.0  & 17.8 & 6.2 \\ 
60 & 0.41 & 0.38 & Crossing 2 & 6.0 & 7.4 & 7.5 & 7.5 & 8.4 & 5.8 \\ 
60 & 0.19 & 0.38 & Delayed Diverging & 29.5 & 37.0 & 37.3 & 35.8 & 37.4 & 31.8 \\ 
60 & 0.17 & 0.38 & Diverging & 40.6 & 45.2 & 45.4 & 42.6  & 41.0 & 39.8 \\ 
60 & 0.29 & 0.38 & Converging 1 & 16.5 & 11.2 & 14.9 & 15.2 & 13.4 & 10.2 \\ 
60 & 0.45 & 0.38 & Converging 2 & 11.6 & 8.4 & 11.0 & 11.0 & 10.8 & 7.1 \\ 
60 & 0.23 & 0.38 & Constant & 27.6 & 24.8 & 26.4 & 24.2 & 21.4 & 22.7 \\ 
120 & 0.35 & 0.38 & Crossing 1 & 5.0 & 10.2 & 13.6 & 22.4  & 30.4 & 7.6 \\ 
120 & 0.41 & 0.38 & Crossing 2 & 5.7 & 8.8 & 8.6 & 9.2 & 11.6 & 7.0 \\ 
120 & 0.19 & 0.38 & Delayed Diverging & 49.1 & 62.2 & 64.3 & 60.7  & 61.6 & 53.7 \\ 
120 & 0.17 & 0.38 & Diverging & 68.0 & 73.4 & 73.9 & 70.0  & 68.2 & 67.1 \\ 
120 & 0.28 & 0.38 & Converging 1 & 24.8 & 18.0 & 23.4 & 24.4 & 22.2 & 16.7 \\ 
120 & 0.45 & 0.38 & Converging 2 & 16.8 & 11.1 & 16.0 & 18.4  & 16.2 & 10.4 \\ 
120 & 0.23 & 0.38 & Constant & 47.0 & 41.7 & 44.8 & 40.9  & 37.6 & 38.3 \\ 
240 & 0.34 & 0.37 & Crossing 1 & 4.8 & 12.8 & 22.2 & 40.9  & 55.4 & 11.2 \\ 
240 & 0.41 & 0.37 & Crossing 2 & 6.1 & 13.8 & 11.2 & 12.6  & 19.4 & 11.4 \\ 
240 & 0.19 & 0.37 & Delayed Diverging & 78.0 & 88.6 & 91.8  & 90.6 & 91.2 & 84.2 \\ 
240 & 0.17 & 0.37 & Diverging & 93.4 & 95.8 & 96.2 & 95.0  & 94.2 & 93.8 \\ 
240 & 0.28 & 0.37 & Converging 1 & 43.4 & 29.0 & 41.9 & 44.2 & 40.6 & 28.6 \\ 
240 & 0.45 & 0.37 & Converging 2 & 31.4 & 17.8 & 29.3 & 33.6 & 30.4 & 18.8 \\ 
240 & 0.23 & 0.37 & Constant & 75.5 & 69.4 & 72.9 & 70.0 & 66.0 & 67.0 \\ 
60 & 0.56 & 0.60 & Crossing 1 & 6.2 & 10.2 & 9.0 & 9.0 & 10.9 & 7.8 \\ 
60 & 0.63 & 0.60 & Crossing 2 & 6.2 & 7.3 & 6.7 & 6.8  & 6.5 & 5.0 \\ 
60 & 0.40 & 0.60 & Delayed Diverging & 31.2 & 40.8 & 38.6 & 36.1  & 33.6 & 35.3 \\ 
60 & 0.38 & 0.60 & Diverging & 39.4 & 43.5 & 42.5 & 40.2 & 36.2 & 37.4 \\ 
60 & 0.52 & 0.60 & Converging 1 & 10.8 & 8.9 & 9.8 & 10.2  & 8.0 & 6.7 \\ 
60 & 0.65 & 0.60 & Converging 2 & 7.4 & 6.0 & 7.0 & 7.8 & 6.6 & 4.9 \\ 
60 & 0.46 & 0.60 & Constant & 20.2 & 16.8 & 18.8 & 18.0 & 14.8 & 14.4 \\ 
120 & 0.56 & 0.60 & Crossing 1 & 6.4 & 13.0 & 11.7 & 11.1 & 17.8 & 10.2 \\ 
120 & 0.63 & 0.60 & Crossing 2 & 6.6 & 7.8 & 7.8 & 7.4 & 7.4 & 6.7 \\ 
120 & 0.40 & 0.60 & Delayed Diverging & 54.7 & 70.6 & 66.3 & 62.2  & 59.8 & 64.4 \\ 
120 & 0.38 & 0.60 & Diverging & 66.5 & 73.6 & 71.3 & 68.8 & 64.0 & 67.9 \\ 
120 & 0.52 & 0.60 & Converging 1 & 14.7 & 9.6 & 13.3 & 13.8  & 12.8 & 9.2 \\ 
120 & 0.66 & 0.60 & Converging 2 & 10.8 & 7.2 & 10.2 & 10.3  & 10.8 & 6.6 \\ 
120 & 0.46 & 0.60 & Constant & 33.7 & 29.0 & 31.9 & 30.6  & 25.8 & 25.8 \\ 
240 & 0.56 & 0.60 & Crossing 1 & 7.2 & 20.8 & 18.2 & 17.6 & 30.6 & 18.0 \\ 
240 & 0.64 & 0.60 & Crossing 2 & 7.0 & 12.6 & 10.2 & 9.1  & 9.6 & 10.8 \\ 
240 & 0.40 & 0.60 & Delayed Diverging & 84.4 & 94.5 & 92.4 & 90.8 & 89.5 & 92.6 \\ 
240 & 0.38 & 0.60 & Diverging & 92.6 & 96.0 & 95.4 & 94.7  & 92.9 & 94.0 \\ 
240 & 0.52 & 0.60 & Converging 1 & 24.4 & 15.2 & 23.0 & 24.2 & 22.0 & 14.3 \\ 
240 & 0.66 & 0.60 & Converging 2 & 17.4 & 9.6 & 15.4 & 17.0  & 17.0 & 9.0 \\ 
240 & 0.46 & 0.60 & Constant & 58.6 & 50.4 & 56.6 & 54.4 & 47.4 & 48.4 \\ 
   \hline
\end{longtable}

To have an overall evaluation of the methods under crossing hazards and also across different alternatives, we borrow the idea of multi-criteria decision analysis and calculate a total score for each method via the following procedure: first rank each test by their power under each scenario, then sum up scores of each test over crossing hazards and all alternatives. The final scores are based on various sample sizes and censoring rates; results are displayed in Table \ref{tab:totscore2} for Type I censoring mechanism and in Table \ref{tab:totscore1} for Type II censoring mechanism. For the Type II censoring design, the proposed test - $\varphi^*(0.5)$ has the highest score, followed by the projection test and Maxcombo test under crossing hazards setting. The Maxcombo test has the highest score if all alternatives are considered, followed by the proposed test. Regarding the Type I censoring mechanism, the total scores across all scenarios have the same ranking as in the Type II censoring. Under the crossing hazard, the projection test has the highest score and then the proposed test. As discussed above, this is mainly because $\theta$ used in the test is 0.5, which is different from the actual value in some scenarios. Even so, the score of the $\varphi^*(0.5)$ test is better than that of the Maxcombo test. 

\begin{longtable}[r]{p{0.06\textwidth}p{0.04\textwidth}p{0.04\textwidth}p{0.16\textwidth}p{0.06\textwidth}p{0.06\textwidth}p{0.06\textwidth}p{0.06\textwidth}p{0.06\textwidth}p{0.06\textwidth}}
\caption{Empirical power under Type II censoring mechanism (\%)}
\label{tab:res2_pwr}\\
  \hline
  Event/N & $\phi_1$ & $\phi_0$ & Type & Logrank & FH11 & maxC & $\varphi^*(0.5)$ & ProjT & Renyi \\ 
  \hline
  \endhead
  \hline   
  \endfoot
 50/60 & 0.18 & 0.15 & Crossing 1 & 19.0 & 13.2 & 23.6 & 30.8 & 30.2 & 17.4 \\ 
 50/60 & 0.16 & 0.17 & Crossing 2 & 10.8 & 6.0 & 13.5 & 19.2  & 20.1 & 6.9 \\ 
 50/60 & 0.13 & 0.20 & Delayed Diverging & 10.0 & 9.4 & 12.2 & 12.0  & 13.4 & 7.8 \\ 
 50/60 & 0.11 & 0.22 & Diverging & 21.4 & 21.2 & 23.5 & 21.4  & 21.8 & 17.2 \\ 
 50/60 & 0.12 & 0.21 & Converging 1 & 30.8 & 23.2 & 29.8 & 29.0  & 27.0 & 22.8 \\ 
 50/60 & 0.20 & 0.13 & Converging 2 & 21.6 & 16.2 & 21.2 & 20.7 & 19.2 & 16.4 \\ 
 50/60 & 0.11 & 0.22 & Constant & 31.6 & 28.0 & 30.1 & 27.1  & 25.8 & 25.4 \\ 
 100/120 & 0.18 & 0.15 & Crossing 1 & 33.3 & 22.6 & 45.2 & 57.8  & 57.0 & 33.8 \\ 
 100/120 & 0.16 & 0.17 & Crossing 2 & 14.0 & 7.4 & 21.6 & 36.8  & 36.8 & 8.2 \\ 
 100/120 & 0.13 & 0.20 & Delayed Diverging & 13.1 & 12.4 & 16.3 & 17.6  & 19.2 & 9.4 \\ 
 100/120 & 0.11 & 0.22 & Diverging & 35.5 & 36.4 & 39.4 & 35.6  & 35.8 & 29.7 \\ 
 100/120 & 0.12 & 0.21 & Converging 1 & 52.1 & 39.4 & 50.5 & 48.4 & 46.2 & 38.3 \\ 
 100/120 & 0.20 & 0.13 & Converging 2 & 39.4 & 29.7 & 39.0 & 37.8  & 35.8 & 29.7 \\ 
 100/120 & 0.11 & 0.22 & Constant & 54.2 & 48.8 & 50.9 & 46.2 & 43.8 & 44.2 \\ 
 200/240 & 0.18 & 0.15 & Crossing 1 & 55.4 & 37.9 & 76.6 & 86.8  & 86.8 & 64.2 \\ 
 200/240 & 0.16 & 0.17 & Crossing 2 & 19.8 & 7.2 & 40.2 & 65.4  & 65.8 & 10.8 \\ 
 200/240 & 0.13 & 0.20 & Delayed Diverging & 18.2 & 15.2 & 26.4 & 29.0  & 32.0 & 11.9 \\ 
 200/240 & 0.11 & 0.22 & Diverging & 61.2 & 61.5 & 67.4 & 63.4  & 63.4 & 52.8 \\ 
 200/240 & 0.12 & 0.21 & Converging 1 & 80.2 & 67.1 & 81.2 & 79.8  & 78.4 & 66.7 \\ 
 200/240 & 0.20 & 0.13 & Converging 2 & 66.0 & 51.8 & 67.0 & 64.4  & 62.4 & 53.3 \\ 
 200/240 & 0.11 & 0.22 & Constant & 81.8 & 77.0 & 79.1 & 75.6  & 73.8 & 73.4 \\
 40/60 & 0.38 & 0.28 & Crossing 1 & 25.1 & 16.9 & 25.3 & 29.4  & 27.8 & 19.4 \\ 
 40/60 & 0.31 & 0.36 & Crossing 2 & 13.0 & 7.0 & 13.9 & 18.6  & 19.7 & 7.2 \\ 
 40/60 & 0.31 & 0.35 & Delayed Diverging & 7.0 & 6.1 & 7.9 & 7.8 & 8.2 & 5.4 \\ 
 40/60 & 0.28 & 0.38 & Diverging & 15.1 & 16.0 & 16.6 & 14.8  & 14.7 & 12.4 \\ 
 40/60 & 0.27 & 0.40 & Converging 1 & 29.6 & 22.2 & 27.1 & 27.1 & 24.6 & 20.9 \\ 
 40/60 & 0.39 & 0.28 & Converging 2 & 21.5 & 15.2 & 19.9 & 21.2  & 18.6 & 14.6 \\ 
 40/60 & 0.27 & 0.40 & Constant & 27.3 & 24.0 & 26.2 & 23.9  & 22.2 & 21.2 \\ 
 80/120 & 0.39 & 0.28 & Crossing 1 & 44.6 & 29.1 & 49.0 & 55.8  & 53.2 & 37.6 \\ 
 80/120 & 0.31 & 0.36 & Crossing 2 & 18.8 & 8.0 & 24.4 & 35.4 & 36.2 & 9.4 \\ 
 80/120 & 0.31 & 0.35 & Delayed Diverging & 7.8 & 8.4 & 9.7 & 9.8 & 10.4 & 6.8 \\ 
 80/120 & 0.29 & 0.38 & Diverging & 24.0 & 25.8 & 25.8 & 22.7  & 22.8 & 21.3 \\ 
 80/120 & 0.27 & 0.40 & Converging 1 & 49.4 & 35.6 & 47.4 & 46.8  & 43.2 & 35.4 \\ 
 80/120 & 0.39 & 0.28 & Converging 2 & 39.4 & 28.1 & 37.1 & 37.4  & 34.0 & 27.8 \\ 
 80/120 & 0.27 & 0.40 & Constant & 44.0 & 39.2 & 42.5 & 38.2  & 36.0 & 36.6 \\ 
  160/240 & 0.39 & 0.28 & Crossing 1 & 72.4 & 51.4 & 80.0 & 85.2  & 83.8 & 69.2 \\ 
 160/240 & 0.31 & 0.36 & Crossing 2 & 31.2 & 9.4 & 43.8 & 65.8 & 64.9 & 15.0 \\ 
 160/240 & 0.31 & 0.35 & Delayed Diverging & 8.2 & 8.6 & 11.4 & 11.2  & 13.0 & 6.4 \\ 
 160/240 & 0.29 & 0.38 & Diverging & 41.8 & 44.6 & 46.6 & 43.0  & 43.0 & 37.3 \\ 
 160/240 & 0.27 & 0.40 & Converging 1 & 78.0 & 61.3 & 77.0 & 76.8  & 74.6 & 61.8 \\ 
 160/240 & 0.39 & 0.28 & Converging 2 & 66.9 & 50.3 & 65.4 & 65.4  & 61.0 & 50.9 \\ 
 160/240 & 0.27 & 0.40 & Constant & 73.8 & 67.4 & 71.2 & 67.0  & 63.4 & 63.2 \\ 
 30/60 & 0.56 & 0.44 & Crossing 1 & 26.1 & 17.8 & 25.4 & 27.2  & 25.4 & 19.3 \\ 
 30/60 & 0.46 & 0.53 & Crossing 2 & 13.8 & 6.9 & 13.6 & 17.2  & 17.0 & 7.2 \\ 
 30/60 & 0.49 & 0.51 & Delayed Diverging & 6.1 & 5.8 & 6.5 & 6.2 & 6.1 & 5.6 \\ 
 30/60 & 0.46 & 0.54 & Diverging & 10.5 & 11.8 & 12.2 & 11.5  & 10.9 & 8.9 \\ 
30/60 & 0.44 & 0.56 & Converging 1 & 26.3 & 17.6 & 23.8 & 23.0  & 20.2 & 16.3 \\ 
 30/60 & 0.56 & 0.44 & Converging 2 & 19.7 & 14.6 & 18.8 & 19.8  & 18.0 & 13.0 \\ 
 30/60 & 0.44 & 0.55 & Constant & 22.0 & 18.4 & 20.8 & 19.6 & 17.2 & 15.2 \\ 
  60/120 & 0.56 & 0.44 & Crossing 1 & 46.5 & 29.3 & 45.9 & 50.3 & 46.0 & 34.4 \\ 
  60/120 & 0.46 & 0.53 & Crossing 2 & 22.2 & 8.6 & 24.1 & 33.0 & 34.4 & 9.8 \\ 
  60/120 & 0.49 & 0.51 & Delayed Diverging & 6.2 & 6.0 & 6.6 & 6.5 & 6.7 & 5.1 \\ 
  60/120 & 0.46 & 0.54 & Diverging & 16.6 & 18.6 & 18.4 & 16.6  & 16.2 & 15.2 \\ 
  60/120 & 0.44 & 0.56 & Converging 1 & 42.4 & 29.2 & 39.5 & 40.7 & 37.3 & 28.2 \\ 
  60/120 & 0.56 & 0.44 & Converging 2 & 35.0 & 23.8 & 33.6 & 33.7 & 30.4 & 23.6 \\ 
  60/120 & 0.44 & 0.55 & Constant & 35.8 & 30.6 & 33.7 & 32.2 & 28.6 & 27.8 \\ 
  120/240 & 0.56 & 0.44 & Crossing 1 & 75.2 & 52.4 & 76.6 & 78.7 & 75.7 & 63.3 \\ 
  120/240 & 0.46 & 0.53 & Crossing 2 & 35.1 & 10.2 & 43.2 & 60.4 & 60.4 & 14.7 \\ 
  120/240 & 0.49 & 0.51 & Delayed Diverging & 6.4 & 6.0 & 6.3 & 6.2 & 7.2 & 5.0 \\ 
  120/240 & 0.46 & 0.54 & Diverging & 27.4 & 31.6 & 31.6 & 28.4  & 27.2 & 26.8 \\ 
  120/240 & 0.44 & 0.56 & Converging 1 & 70.7 & 51.4 & 69.1 & 68.9 & 65.0 & 50.6 \\ 
  120/240 & 0.56 & 0.44 & Converging 2 & 60.6 & 44.8 & 58.3 & 58.2 & 54.0 & 43.9 \\ 
  120/240 & 0.44 & 0.55 & Constant & 60.8 & 53.6 & 57.4 & 54.4  & 50.5 & 50.6 \\ 
     \hline
\end{longtable}


\begin{table}[ht]
\centering
\caption{Ranking scores of tests under Type I censoring mechanism}
\label{tab:totscore2}
\begin{tabular}{rrrrrrr}
  \hline
 & Logrank & FH11 & maxC & $\varphi^*(0.5)$ & ProjT & Renyi \\ 
  \hline
Crossing & 31 & 62 & 70 & 78 & 99 & 38 \\ 
  Total & 220 & 202 & 283 & 272 & 236 & 110 \\ 
   \hline
\end{tabular}
\end{table}

\begin{table}[ht]
\centering
\caption{Ranking scores of tests under Type II censoring mechanism}
\label{tab:totscore1}
\begin{tabular}{rrrrrrr}
  \hline
 & Logrank & FH11 & maxC & $\varphi^*(0.5)$ & ProjT & Renyi \\ 
  \hline
Crossing & 57 & 18 & 70 & 102 & 93 & 38 \\ 
  Total & 270 & 144 & 296 & 280 & 242 & 90 \\ 
   \hline
\end{tabular}
\end{table}


\subsection{Sensitivity of the test to the crossing time point}
We recommend the default value 0.5 for $\theta$ if the information about when the crossing occurs is not available. In this section, we will investigate whether the performance of the test is sensitive to the choice of $\theta$ through simulation. The six crossing scenarios for sample size 240 under the Type I censoring mechanism are assumed. The empirical power of the proposed test with $\theta$ varying from 0.1 to 0.9 is simulated and shown in Table \ref{tab:sens}. We can see that the power varies a bit with different values of $\theta$. For example, with $\phi_0=0.60$ and crossing 1, the highest power is 34.3\% at $\theta=0.1$ and the lowest power occurs at $\theta=0.6$. In this case, the crossing occurs at the very beginning, and if the researcher mistakenly specifies 0.9 for $\theta$, the power is 17.7\%, which is larger than the power of logrank test (7.2\%) and close to the power of the Maxcombo test (18.2\%). In general, the value of $\theta$ has some impacts on the power of the test, but even in the worst case, it can achieve similar power as the Maxcombo test and higher power than the logrank test. 
\begin{table}[ht!]
\centering
\caption{Empirical power of the proposed tests at different $\theta$s}
\label{tab:sens}
\begin{tabular}{ccccrrrrrrrrrr}
  \hline
  N & $\phi_1$ & $\phi_0$ & Type & \multicolumn{9}{c}{$\theta$}\\ 
  &&&& 0.1 & 0.2 & 0.3 & 0.4 & 0.5 & 0.6 & 0.7 & 0.8 & 0.9 \\ 
  \hline
240 & 0.18 & 0.18 & crossing 1 & 58.70 & 66.40 & 74.40 & 79.20 & 81.10 & 79.00 & 73.00 & 66.00 & 59.80 \\ 
 240 & 0.19 & 0.18 & crossing 2 & 33.90 & 45.40 & 48.90 & 47.70 & 43.00 & 37.90 & 31.80 & 27.60 & 23.80 \\ 
240 & 0.34 & 0.37 & crossing 1 & 40.80 & 54.00 & 55.40 & 50.40 & 40.90 & 34.10 & 28.40 & 25.40 & 24.00 \\ 
240 & 0.41 & 0.37 & crossing 2 & 23.60 & 23.30 & 19.10 & 15.20 & 12.60 & 11.40 & 11.00 & 10.70 & 10.60 \\ 
240 & 0.56 & 0.60 & crossing 1 & 34.30 & 31.20 & 23.30 & 19.10 & 17.50 & 17.40 & 17.60 & 17.60 & 17.70 \\ 
 240 & 0.64 & 0.60 & crossing 2 & 13.20 & 10.40 & 8.60 & 9.00 & 9.00 & 9.10 & 9.40 & 9.40 & 9.60 \\ 
   \hline
\end{tabular}
\end{table}
\subsection{One-sided test}
Figure ~\ref{fig:example1 } is created from Type I censoring mechanism with $\alpha=2,\beta=9$. It illustrates the stochastic hazards. Suppose the alternative hypothesis of interest is $H_a: \Lambda_1 \geq \Lambda_0$. The power of the one-sided test with significance level 0.025 is simulated for the logrank test, Maxcombo test, and our proposed $\varphi^*(0.5)$ test. The censoring rates for Group 1 and Group 0 are 7.4\% and 7\%. At sample size 120, the corresponding power values are 26.6\%, 42.6\%, and 53.3\% for the three tests, respectively. The Maxcombo test and the proposed test both have more power than the logrank test.  
\section{Real Data Application}
\begin{figure}
    \centering
  
     \includegraphics[scale=0.38]{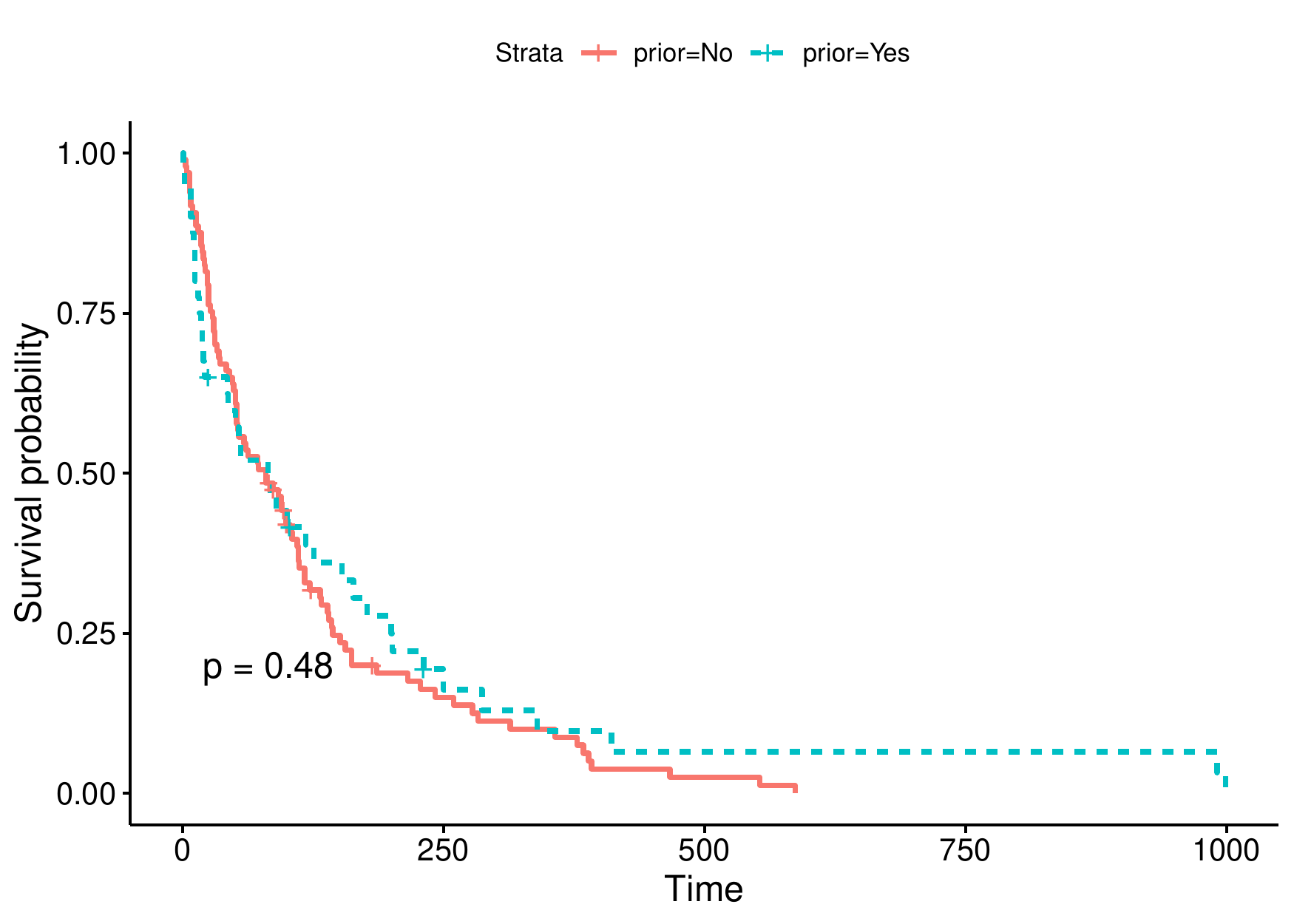}
       \includegraphics[scale=0.38]{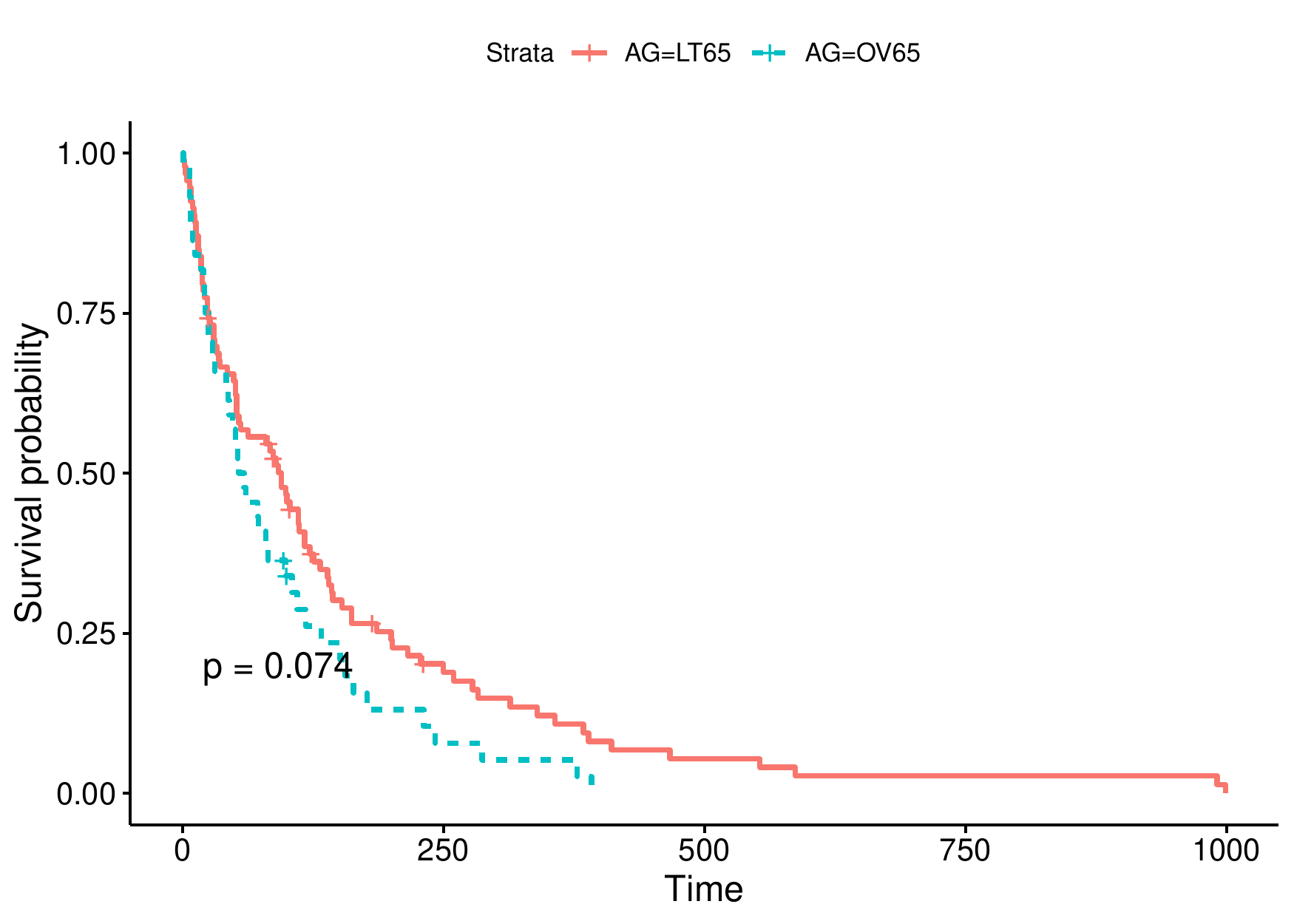}
    \caption{KM plots for Case 1 (left) and Case 2 (right)}
    \label{fig:realdata}
\end{figure}

The real data is from appendix A of \cite{kalbfleisch2011statistical}. The data is about a randomized trial of two treatments for lung cancer. Besides treatment-related information, several covariates were collected. In this example, we use two covariates whether the patient had prior therapies and whether the patient was older than 65 years. A total of ninety-seven patients did not receive prior therapies, and forty subjects received therapies. Ninety-three patients are younger than 65, and forty-four patients are older than 65. The Kaplan-Meier curves by prior therapy and age are given in Figure \ref{fig:realdata}. The event rates are all beyond 90\% in each group. On the left plot (prior therapy), the two curves cross when the survival probabilities are around 50\%. Thus, patients receiving prior therapies have a lower survival rate initially but survive longer in the long run. The survival plot on the right (Age>65) is more likely to be from a process satisfying the proportional hazard. Seven competing tests, including logrank, Renyi, Maxcombo, projection, and the proposed method with $\theta=0.25, 0.5,0.75$ are applied to this real data. 

The p-values of all tests are shown in Table \ref{tab: realdata}. In the prior therapy group, the logrank test gives a p-value equal to 0.48. All the rest tests have smaller p-values and the smallest one is from the proposed method with $\theta=0.25$. This is because the survival curves cross at around 0.5, indicating the hazard must cross before 0.5. A correct specification of $\theta$ increases the power. Even if the incorrect selection of $\theta=0.75$ yields a p-value close to the one given by the Maxcombo test. It shows that the proposed method gains a lot in the correct specification of $\theta$ but loses a little with an incorrect choice in the crossing hazard scenario. The projection test also gives a reasonably small p-value. In the age group, the logrank test has the smallest p-value of 0.07, showing the optimality of the test under proportional hazard. Renyi's test and projection test have the largest p-values, and the Maxcombo and proposed test have close p-values in between.  These results are consistent with the simulation. The projection test loses more power than the Maxcombo test under proportional hazard. 

\begin{table}[ht]
\centering

\caption{P-values of tests for the lung cancer data}
\label{tab: realdata}
\begin{tabular}{rrrrrrrr}
  \hline
 & Logrank & Renyi & maxC & ProjT & $\varphi^*(0.25)$ & $\varphi^*(0.5)$ & $\varphi^*(0.75)$ \\ 
  \hline
Prior therapy & 0.48 & 0.38 & 0.28 & 0.19 & 0.10 & 0.24 & 0.3 \\ 
  Age>65 & 0.07 & 0.16 & 0.10 & 0.14 & 0.12 & 0.12 & 0.10 \\ 
   \hline
\end{tabular}
\end{table}

\section{Discussions }
In real-world studies, the proportional hazards assumption is predominately assumed in analyzing time-to-event data, even though the assumption may not hold in two distinct situations (\cite{saad2018understanding}). One is that the treatment effects interact with patient characteristics; the other is that the treatment effects vary with time. Both cases are not rare in oncology trials.  Therefore, researchers are motivated to find an omnibus test that performs well in most situations. In this study, we have proposed a maximum weighted logrank test, which particularly incorporates the weight for detecting crossing hazards. Through simulations under various sample sizes and censoring rates, we show that our proposed test has a sizeable gain in power under crossing hazards regardless of the selection of nuisance parameter $\theta$ compared to the logrank test. At the worst selection of $\theta$, the power is comparable to the Maxcombo test, while the power increases noticeably if $\theta$ approximates the true crossing point.  For the converging and proportional hazards, similar to the Maxcombo test, the proposed test has some power loss compared to the logrank test. The projection test is a comparative method in detecting crossing hazards, but it loses more power than the proposed test at the proportional and converging hazards. 

If there is prior information suggesting the crossing hazards may hold, we can extend the proposed test to one that is only sensitive to the crossing hazards. For example, if we let $g_{\theta}(u)$ denote the proposed crossing weight function, that is, $g_{\theta}(u)=\mathbb{1}(u\leq \theta)\frac{u-\theta}{\theta}+\mathbb{1}(u> \theta)\frac{u-\theta}{1-\theta}$. The test denoted by $\varphi^*(0.2,0.5,0.8)$ with weights - $\Tilde{w}_1=1,\Tilde{w}_2=g_{1/5}(u),\Tilde{w}_3=g_{1/2}(u),\Tilde{w}_4=g_{4/5}(u),\; u \in (0,1)$ may further improve the power of the proposed test under crossing hazards. This test addresses early, middle, and late crossing scenarios, so there is no need to guess the actual crossing point. The same simulation scenarios as described in section 4 are used. The empirical power and Type I error rates based on the Type I censoring mechanism are shown in Table~\ref{sensonly_t}. Cases with $\beta=15,25,40$ correspond to small, medium and large censoring rates, same as the numbers shown in Table \ref{tab:fres2}. If we compare the power with the projection tests under Crossing 1 alternatives, this maximum test shows power advantage in most cases. For example, with N=240, $\beta=25$ and Crossing 1 alternative, respectively, the power for the logrank, Maxcombo, proposed - $\varphi^*(0.5)$ and projection test are 4.8\%, 22.2\%, 40.9\% and 55.4\%. The power for this maximum test is 56.7\%, larger than all the previous tests. However, the power loss under proportional hazard is greater than the Maxcombo test and proposed test-$\varphi^*(0.5)$. For example, with N=240, $\beta=40$ and constant hazard respectively, the power for logrank, Maxcombo, $\varphi^*(0.5)$ and projection test are 58.6\%, 56.6\%, 54.4\% and 47.4\%. The power for this maximum test is 49.7\%, only slightly larger than the power of the projection test.  The Type I error rates are similar to the Maxcombo test and proposed test-$\varphi^*(0.5)$. 
\begin{table}[ht]
\centering
\caption{Empirical power and type I error rates}
\label{sensonly_t}
\begin{tabular}{rllllllllll}
  \hline
 & Hazards &  \multicolumn{3}{c}{$\beta=15$}&\multicolumn{3}{c}{$\beta=25$}&\multicolumn{3}{c}{$\beta=40$}\\&&N=60 & N=120 & N=240 & N=60 & N=120 & N=240 & N=60& N=120 & N=240 \\ 
  \hline
 & Crossing 1 & 0.267 & 0.485 & 0.814 & 0.173 & 0.322 & 0.567 & 0.096 & 0.158 & 0.271 \\ 
   & Constant & 0.251 & 0.437 & 0.756 & 0.212 & 0.373 & 0.66 & 0.158 & 0.261 & 0.497 \\ 
   & $H_0$ & 0.062 & 0.057 & 0.046 & 0.06 & 0.054 & 0.046 & 0.052 & 0.05 & 0.048 \\ 
   \hline
\end{tabular}
\end{table}

In summary, we recommend to use $\varphi^*(0.2,0.5,0.8)$ test or projection test if  crossing hazards scenario is the most likely one. However, if there is not enough prior information about the alternative hazards, the proposed $\varphi^*(0.5)$ test and Maxcombo test are more appropriate as they have quite robust power gain in various non-proportional hazards. We have complied all the functions used in  the simulation including the maximum logrank test and projection test in an R package on GitHub (hcheng1118/maxLRT). Considering the various advantages of the maximum logrank test, we are also working on a project that considers different scenarios of non-proportional hazards in the design phase and proposes a simulation-free sample size calculation procedure based on the proposed test. The manuscript is to be finished soon. Applying the method in the adaptive design is also our future work.

\printendnotes
\clearpage 
\bibliography{sample.bib}

\end{document}